# A Privacy-Preserving Recommender for Filling Web Forms Using a Local Large Language Model

Amirreza Nayyeri
Department of Computer Engineering
Faculty of Engineering
Ferdowsi University of Mashhad
nayyeri.amirreza@mail.um.ac.ir

Abbas Rasoolzadegan
Department of Computer Engineering
Faculty of Engineering
Ferdowsi University of Mashhad
rasoolzadegan@um.ac.ir

*Abstract*— Web applications are increasingly used in critical domains such as education, finance, and e-commerce. This highlights the need to ensure their failure-free performance. One effective method for evaluating failure-free performance is web form testing, where defining effective test scenarios is key to a complete and accurate evaluation. A core aspect of this process involves filling form fields with suitable values to create effective test cases. However, manually generating these values is time-consuming and prone to errors. To address this, various tools have been developed to assist testers. With the appearance of large language models (LLMs), a new generation of tools seeks to handle this task more intelligently. Although many LLM-based tools have been introduced, as these models typically rely on cloud infrastructure, their use in testing confidential web forms raises concerns about unintended data leakage and breaches of confidentiality. This paper introduces a privacy-preserving recommender that operates locally using a large language model. The tool assists testers in web form testing by suggesting effective field values. This tool analyzes the HTML structure of forms, detects input types, and extracts constraints based on each field's type and contextual content, guiding proper field filling. A comparative evaluation with the T5-GPT approach on five web apps shows comparable performance (63 vs. 64 detected forms), with our local approach excelling in privacy preservation by avoiding external dependencies. Real-world testing on ten Persian-language websites yields high metrics: 92.9% accuracy, 94.4% precision, and 98% recall in field detection and value generation. These results demonstrate that the proposed approach delivers high accuracy in input generation and yields fewer failures. Overall, the study confirms the tool's effectiveness in locally filling web forms while preserving privacy, with robust performance across diverse, real-world UI contexts, including non-English languages.

## I. INTRODUCTION

The rapid evolution of web-based technologies has led to the unavoidable integration of web applications into daily life and contemporary business processes [1]. Web applications are widely used in various fields, including education, medicine, finance, and e-commerce. The extensive use of web applications has significantly increased the demand for ensuring they operate without failure. Any such failure can have far-reaching consequences for organizations and individuals, undermining reputations, inflicting financial losses, and draining valuable resources [2]. In software quality, this is referred to as reliability, which is the ability of software to perform its required functions under stated conditions for a specified period [3].

Software reliability is difficult to evaluate due to software complexity [4]. One of the most effective strategies to evaluate software reliability is software testing, which enables the detection of failures before deployment [5]. Graphical user interface (GUI)-based black-box testing, exemplified by tools, remains one of the most widely used and effective methods for evaluating web and mobile application functionality. Its strength lies in applicability without requiring implementation details, making it both popular and successful in practice [6].

In black-box testing, defining test scenarios plays a crucial role [7]. A test scenario outlines the "what to test" but must then be complemented with effective data and context to form a comprehensive test case. Selecting effective test values ensures that each test is realistic, repeatable, and comparable, increasing coverage—especially for edge cases—and minimizing human error [8].

Recent advances in Large Language Models (LLMs), using models such as Chat-GPT, Gemini, and LLaMA, have made them powerful candidates for implementing such a recommender. These models are capable of understanding form contexts, generating test scripts, and even identifying user interface (UI) components [9]. Their ability to produce human-like and contextually relevant outputs makes them particularly well-suited for supporting testers. However, challenges such as requirements for privacy preservation associated with cloud-based AI services [10] still exist. When confidential source codes must be tested, sending inputs to external servers introduces potential risks of data leakage or policy violations. Running such recommenders locally ensures privacy-preserving operation.

This paper presents a local, LLM-powered privacy-preserving recommender designed to assist testers by generating effective values for web form fields. The proposed method generates structured, context-aware test inputs for each form field. In the method, it fills forms using the first valid suggestion. Designed to run locally, the proposed method is tailored for confidential web forms, ensuring privacy-preserving operation.

The primary goal of this study is to develop a local privacy-preserving recommender that supports testers in generating effective values for web forms for rapid testing and higher accuracy. Our approach demonstrates how a local LLM-powered recommender can be integrated into real-world testing workflows without compromising requirements for confidentiality or accuracy.

The rest of this paper is organized as follows. Section II reviews relevant work on testing, with a focus on the use of LLMs. Section III describes the proposed method and its components, including the browser extension and local model. Section IV presents experimental results and evaluation metrics. Finally, Section V concludes the paper with insights and future directions.

## II. RELATED WORK

In recent years, form input generation has garnered notable attention due to the increasing complexity of web applications. As mentioned before, generating effective form inputs is a key part of test scenario definition, as it directly impacts the reliability of web application testing. In this section, related studies are briefly introduced in three categories: crawling-based techniques, optimization-based methods, and large language model (LLM)-based solutions. Each category advances by filling out web forms, but they often overlook aspects such as privacy preservation. This section reviews these approaches, with a focus on their form-filling capabilities, and highlights the gaps that the proposed method addresses.

### A. Crawling-Based Techniques

Crawling-based techniques aim to exhaustively explore web applications by constructing a model of navigable states. These methods typically rely on user interactions or heuristic rules to fill forms during the crawling process.

Sunman et al. [11] record user interactions during exploratory testing to build a crawl graph that prioritizes unexplored pages. This approach leverages user-provided inputs for form filling, ensuring that the inputs are realistic and relevant to the application. However, it requires initial manual effort to record the interactions, and the coverage of form inputs is limited to what the user provides, potentially missing edge cases such as invalid inputs.

Similarly, Lin et al. [12] augment traditional crawling with semantic-aware heuristics to focus on high-impact paths. By incorporating semantic information from web pages, they aim to generate more meaningful inputs for form fields, thereby improving the effectiveness of the testing process. Nevertheless, these methods may struggle with highly dynamic forms or those requiring context-specific inputs, as they lack a deep understanding of the semantics of the form fields.

While crawling-based techniques can achieve good coverage of web application states, their form-filling capabilities are often limited by the need for manual input or simplistic heuristic rules, which may not fully explore the range of possible inputs.

### B. Optimization-Based Methods

Optimization-based methods treat UI exploration and form input generation as search problems, employing metaheuristics or learning agents to optimize the process. These approaches aim to generate inputs that systematically maximize coverage or other testing objectives.

Groce [13] uses adaptation-based programming to synthesize input sequences without requiring a complete model of the application. This allows for flexible exploration of the input space, potentially generating diverse form inputs. Carino and Andrews [14] leverage ant-colony optimization to discover effective GUI event sequences, which can include form-filling actions. Their method focuses on optimizing the sequence of interactions to achieve higher coverage.

Zheng et al. [15] propose curiosity-driven reinforcement learning agents to guide exploration. By rewarding the agent for discovering new states, this approach can lead to more efficient test case discovery, including diverse forms of input. However, while these methods can systematically increase coverage, they often lack a deep understanding of the semantics behind form fields. As a result, the generated inputs may be syntactically correct but semantically invalid, reducing their effectiveness in testing real-world scenarios.

Moreover, optimization-based methods typically require significant computational resources, making them less practical for large-scale web applications. Their reliance on search algorithms.

### C. LLM-Based Solutions

LLM-based solutions have emerged as a promising avenue for generating context-aware and realistic inputs for web forms. These methods utilize large language models to comprehend the semantics of form fields and generate suitable inputs.

Nguyen and Maag [16] combine a machine learning (ML) approach to enable codeless testing, which may involve using ML models to predict suitable inputs for form fields. Liu et al. [17] introduce semantic-aware strategies for mobile GUI input generation, which could be adapted for web forms.

In a significant advancement, Chen et al. [18] developed a novel approach for automated web application testing by integrating large language models (LLMs) with data faker libraries to generate diverse and contextually appropriate inputs for web forms. Their method builds upon their reinforcement learning-based model, mUSAGI, addressing its limitations in input diversity, training time, and form submission accuracy. The authors utilized Google's T5 model [19], adapted through prompt tuning, to classify form fields into categories such as email, phone number, or name, using data from 75 webpages across 20 popular websites. Prompt tuning freezes the T5 model's parameters and trains a small model in front, reducing computational costs compared to full model fine-tuning. The Mocker library [20], an open-source tool, then generates realistic input values based on these categories—for example, producing valid email addresses or phone numbers—enhancing the variety and relevance of test inputs. The Mocker library, specifically the mocker-data-generator tool, simplifies testing by creating large amounts of fake but realistic data (e.g., names, addresses) using schema-based generators. This ensures comprehensive testing of web forms, making the process more efficient and effective for identifying defects in web applications. Additionally, GPT-4o [30] is employed to accurately identify submit buttons, overcoming the previous model's reliance on the document object model (DOM) similarity, which often misclassified submissions due to small confirmation texts. This approach achieved a 2.3% increase in statement coverage compared to mUSAGI and 7.7% to 11.9% higher coverage than QExplore [29], a Q-learning-based tool, demonstrating improved test thoroughness. However, the method faces challenges, including privacy concerns due to potential data transmission to external servers for GPT-4o processing and the need for internet access, which limits offline use.

However, despite their high accuracy, LLM-based methods typically rely on cloud-based models, which necessitate internet access and may raise privacy concerns, as sensitive data could be transmitted to external servers. Other LLM-based works, such as Le et al. [21] and Li et al. [22], similarly depend on online services.

### D. Research Gaps

Table compares selected approaches based on key criteria: LLM usage, offline capability, and privacy preservation. LLM-based frameworks rely on online services and neglect privacy concerns. In contrast, the proposed approach provides a local recommender support for generating effective inputs for web forms, ensuring privacy, and enabling offline use. Employing a locally deployed model eliminates the need for external application programming interface (API) calls, thus preserving user privacy and enabling operation in offline environments.

TABLE I: COMPARISON OF SELECTED APPROACHES

| Study | Uses LLM | Offline | Privacy |
|---|---|---|---|
| Sunman et al. [11] | ✗ | ✓ | ✓ |
| Lin et al. [12] | ✗ | ✓ | ✓ |
| Groce [13] | ✗ | ✓ | ✓ |
| Carino & Andrews [14] | ✗ | ✓ | ✓ |
| Zheng et al. [15] | ✗ | ✓ | ✓ |
| Nguyen & Maag [16] | ✓ | ✗ | ✗ |
| Liu et al. [17] | ✓ | ✗ | ✗ |
| Chen et al. [18] | ✓ | ✗ | ✗ |
| Le et al. [21] | ✓ | ✗ | ✗ |
| Li et al. [22] | ✓ | ✗ | ✗ |
| **Our Method** | ✗ | ✓ | ✓ |

In summary, while existing methods have made notable strides in filling web forms, particularly with the integration of LLMs for semantic understanding, they often overlook privacy. The proposed solution addresses these limitations by offering a privacy-aware, and offline-capable approach to form input generation.

## III. THE PROPOSED METHOD

This study introduces a recommender powered by a large language model (LLM) to generate effective values for filling privacy-preserving web forms. The method leverages to perform static HTML analysis to ensure effective value generation. To illustrate the process, we use the registration form from the Persian software website `soft98.ir` as a running example, which includes fields for `username`, `email`, `password`, and `password confirmation`, each governed by specific validation rules.

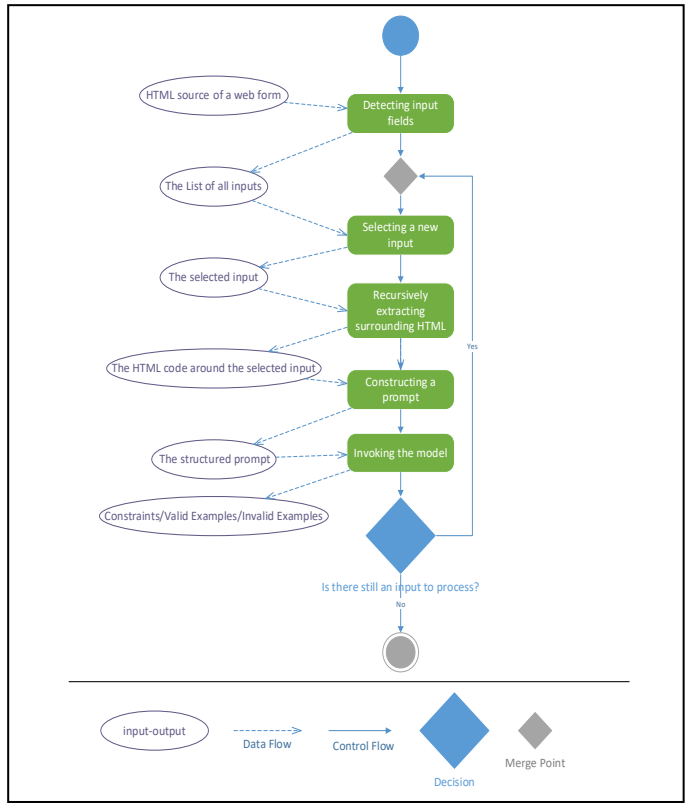

Fig. 1. The process of obtaining initial responses

In this study, we selected the LLaMA 3.1 8B [23] model for our experiments due to its widespread adoption in large language model (LLM)-assisted testing [9] and its open-source nature, which facilitates research and collaboration. The model comprises 8 billion parameters, striking a balance between computational efficiency and performance. This parameter size allows for deployment on systems with moderate computational resources, making it accessible to a broader range of researchers and practitioners.

Furthermore, Llama 3.1 8B boasts a 128K token limit, which means that the input text, when translated into tokens understandable by the model, must not exceed this 128 K limit. This high token limit is particularly advantageous for processing extensive inputs such as HTML web forms. This capability ensures that the model can handle and analyse large prompts without truncation, preserving the integrity of the input data. Being open-source, Llama 3.1 8B promotes transparency and reproducibility in research. It allows researchers to inspect, modify, and distribute the model, fostering innovation and collaboration within the scientific community.

Given these attributes, LLaMA 3.1 8B is well-suited for our research objectives, which involve generating effective input values for web forms. Its combination of computational efficiency, high token limit, and open-source availability aligns perfectly with the requirements of our study.

The method focuses on generating initial values for each input field in a web form. The process is visualized in the activity diagram shown in Figure 1. The method will be explained in 5 steps below.

*Step 1: Detect Input Fields*

The method captures the entire HTML source of the current web page and preprocesses the HTML to identify all visible form fields, specifically targeting `<input>` and `<textarea>` elements for further analysis and example generation.

*Example:* For the `soft98.ir` registration form, the method captures the HTML, which identifies four fields: `username`, `email`, `password`, and `password confirmation`.

*Step 2: Iterate Until All Fields Are Processed*

The method processes each identified input field sequentially. For each unprocessed field, Steps 3 through 5 are executed. Upon completion, the generated outputs are returned to the tester interface and displayed alongside their corresponding fields.

*Example:* The method iterates over the `username`, `email`, `password`, and `password confirmation` fields of the `soft98.ir` form, processing each in turn.

*Step 3: Extract Surrounding HTML Context*

Due to the token limitations of LLMs, such as the LLaMA 3.1 8B [23] model used in this study, which has a limitation of 128k input tokens, the method extracts a relevant portion of the HTML surrounding the target input element. Starting from the input node, it recursively ascends DOM tree, collecting parent elements until the token limit is reached. This ensures that sufficient contextual information is provided without exceeding the model's capacity.

*Example:* For the `password` field in the `soft98.ir` form, the method extracts the input element `<input type="password" name="password" id="password">`, its parent `<div>`, the associated `<label>` ("Password:"), and any nearby validation messages or placeholders, up to the token limit.

*Step 4: Construct a Prompt*

Using the extracted HTML context, the method constructs a structured prompt for the LLM. The prompt instructs the model to act as an HTML parser and generate a JSON object with the following keys: name, id, type, constraints, examples, and bad_examples. The constraints are derived from attributes such as minlength, maxlength, pattern, or placeholder and are described in complete English sentences. The prompt ensures that only the specified element is processed.

*Example:* For the `password` field, the prompt is:

```
For the "password" element, create a JSON
object with keys: name, id, type,
constraints, examples, and bad_examples.
```

- name: The value of the 'name' attribute.
- id: The value of the 'id' attribute (use name if id is absent).
- type: Input type (e.g., text, password) or 'textarea'.
- constraints: Validation rules extracted from attributes like minlength, maxlength, pattern, or placeholder, written in English as complete sentences.
- examples: Five example values that satisfy these constraints and would be accepted by the field's validation.
- bad_examples: Five example values that violate these constraints and would be rejected (for negative testing).

Process only the element with id "password".

*Step 5: Invoke the Model and Retrieve Outputs*

The constructed prompt is sent to LLM, which generates a JSON object containing the requested information. The output is stored for presentation to the tester.

*Example:* For the `password` field, the LLM produces:

```
{
  "name": "password",
  "id": "password",
  "type": "password",
  "constraints": "The password must be at
↪   least 8 characters long.",
  "examples": ["SecurePass123",
↪      "Test!2024", "AnotherValid1",
       "Password!23", "ExamplePass9"],
  "bad_examples": ["1234567", "pass",
↪   "abc", "short", "weak"]
}
```

After obtaining initial responses, the method selects the first valid example from the examples list for each field and fills the form automatically, enabling rapid testing.

## IV. EXPERIMENTAL RESULTS

This section evaluates the proposed method through a comparative analysis with an existing approach on standard web applications, followed by real-world testing on diverse websites. The comparison with T5-GPT approach from Chen et al. [18] demonstrates comparable performance in form detection, while our method's local operation provides a significant advantage in privacy preservation. Overall, this section highlights the method's effectiveness, robustness, and practical advantages in both controlled and real-world scenarios.

Based on the related works discussed, we initially considered several studies for comparison with our research, which focuses on generating effective input values for web forms. However, a closer analysis revealed significant differences in objectives and scope that limit direct comparability with most of these works. Specifically, studies by Sunman et al. [11], Lin et al. [12], Groce [13], Carino & Andrews [14], and Zheng et al. [15] aim to define appropriate test cases to achieve high coverage across HTML elements or entire web forms. This contrasts with our objective, which centers on creating effective input values for each web form, without considering other elements present on the page. Liu et al. [17] designed their work for mobile applications, rendering it challenging to compare with our web-based approach. Li et al. [22] reported numerical accuracies but emphasized comparisons between models rather than the generation of input values or the validation of web forms, making their goals misaligned with ours. Among the potentially comparable studies, the works of Le et al [21], Nguyen & Maag [16], and Chen et al. [18] emerged as relevant candidates. However, Le et al. [21] and Nguyen & Maag [16] do not provide precise details about their datasets, which prevents a meaningful comparison. Consequently, we selected Chen et al. [18] as the most appropriate study for comparison, given its shared focus on generating input values for web forms.

In this section, we proceed with a detailed comparison of our approach and results with those of Chen et al. [18]. For this purpose, we selected Chen et al.'s best-performing approach, T5-GPT, for comparison. T5-GPT is the proposed model in Chen et al's work, which replaces the reinforcement learning agent from their prior mUSAGI model with a large language model (LLM)-based system. Specifically, it uses Google's T5 model with prompt tuning to classify form field categories, a data faker (Mocker) to generate diverse input values based on those categories, and GPT-4o to assist in determining successful form submissions and identifying submit buttons, addressing limitations like fixed input data and imprecise submission detection. We chose T5-GPT for testing because it represents the state-of-the-art in their study, offers a direct alignment with our focus on input value generation for web forms, and allows for a fair evaluation on shared metrics such as form detection, while highlighting our method's unique strengths like local operation without external dependencies.

The method was tested on the same dataset used by Chen et al., focusing on the "input pages" metric—the number of detected forms—as it aligns closely with our method's emphasis on form detection and input generation. Unlike T5-GPT, which incorporates page navigation, our method prioritizes form-specific functionality, making "input pages" the most suitable metric for this comparison.

TABLE II: COMPARISON OF INPUT PAGES DETECTED

| Web App | Proposed Method | T5-GPT |
|---|---|---|
| TimeOff [24] | 15 | 14 |
| NodeBB [25] | 7 | 7 |
| KeystoneJS [26] | 18 | 20 |
| Django [27] | 6 | 6 |
| Petclinic [28] | 17 | 17 |
| **Total Detected** | **63** | **64** |

The proposed method detected one additional input page for TimeOff [24] (15 vs. 14) but two fewer for KeystoneJS [26] (18 vs. 20), matching T5-GPT exactly for NodeBB [25], Django [27], and Petclinic [28]. These results demonstrate comparable performance, with T5-GPT slightly outperforming in total detections (64 vs. 63). The reason for detecting two fewer pages in KeystoneJS is due to one input in each of two forms being mishandled (e.g., incorrect value generation or format mismatch), leading to failed submissions and preventing those pages from being fully explored. However, our method's local operation provides a significant advantage over T5-GPT's reliance on external services, enhancing privacy preservation—a critical factor in sensitive applications. The minor differences in detection may arise from our method's optimization for local efficiency rather than exhaustive navigation, unlike T5-GPT.

Following the comparative analysis, the method was employed to test it on ten Persian-language websites, including e-commerce, educational, and service platforms, as listed in TABLE III. This evaluation focused on form field detection accuracy and the quality of generated inputs, highlighting the method's performance in a non-English language context.

The evaluation results are detailed in TABLE III. The table includes the following columns:

- **Fields**: The total number of input fields present in the website's forms.

- **Correct**: The number of input fields correctly detected as inputs with appropriate examples.

- **Missed**: The number of input fields that the method failed to detect as inputs.

- **Incorrectly Detected**: The number of HTML elements mistakenly identified as input fields by the method.

- **Suboptimal**: The number of input fields where the automatically generated values were incorrect or inappropriate (e.g., wrong format, irrelevant content, or failed validation).

TABLE III: FORM FIELD DETECTION AND CORRECTION BREAKDOWN

| Website | Fields | Correct | Missed | Incorrectly Detected | Suboptimal |
|---|---|---|---|---|---|
| Divar [31] | 15 | 14 | 0 | 0 | 1 |
| Zoomit [32] | 20 | 18 | 2 | 0 | 0 |
| Blog.ir [33] | 20 | 20 | 0 | 0 | 0 |
| Blogfa [34] | 18 | 17 | 0 | 0 | 1 |
| Soft98 [35] | 12 | 12 | 0 | 2 | 0 |
| Digikala [36] | 15 | 14 | 1 | 0 | 0 |
| Ninisite [37] | 22 | 20 | 0 | 2 | 2 |
| Shatel Panel [38] | 10 | 10 | 0 | 0 | 0 |
| Bank Melli Panel [39] | 22 | 20 | 0 | 3 | 2 |
| FUM Portal [40] | 10 | 7 | 0 | 0 | 3 |
| **Total** | **164** | **152** | **3** | **7** | **9** |

In this part, the proposed method is evaluated by several criteria, including: 1) True Positive (TP): The number of fields that were correctly identified as inputs and had appropriate examples (Correct). 2) True Negative (TN): The number of fields that were not input but were incorrectly identified as such (Incorrectly Detected). 3) False Positive (FP): The number of fields that were correctly identified as inputs but had incorrect or inappropriate generated values (Suboptimal). 4) False Negative (FN): The number of fields that were inputs but were

not detected as such (Missed). The results of the proposed method are evaluated with the following criteria:

- **Accuracy**: The proportion of correctly classified fields calculated as $\frac{TP+TN}{TP+FP+TN+FN}$
- **Precision**: The proportion of correctly filled fields among those identified as inputs, calculated as $\frac{TP}{TP+FP}$
- **Recall**: The proportion of correctly filled fields among all actual input fields, calculated as $\frac{TP}{TP+FN}$

Based on Table III, the method identified 152 fields with appropriate examples (TP), 9 fields had incorrect or inappropriate generated values (FP), 7 fields were incorrectly detected as inputs (TN), and 3 fields were missed (FN). The calculated metrics are:

- **Accuracy**: 92.9%
- **Precision**: 94.4%
- **Recall**: 98%

These results indicate high reliability in a Persian-language context, with only three fields missed and seven incorrectly detected, primarily in complex DOM structures or non-standard input wrappers (e.g., Soft98.ir, Ninisite.com, Bank Melli Panel). The nine suboptimal generations reflect instances where the model produced incorrect values (e.g., non-Persian text or invalid formats), underscoring challenges in language-specific handling but demonstrating overall robustness in fully automated operation across diverse, real-world UI contexts.

## V. Conclusion and Future Work

In this paper, we presented a privacy-preserving recommender for filling web forms using a locally deployed large language model. The method analyzes HTML structure and field constraints to generate valid and invalid input examples. In evaluation, our method achieved comparable input page coverage to T5-GPT while preserving user privacy through local execution. In testing on ten Persian websites, the tool reached 92.9% accuracy across 164 fields, confirming its effectiveness in generating effective values. These results demonstrate the effective value of our approach in confidential web forms.

In future work, we aim to scale this method to support testers with a semi-automated mode to provide more accurate results. While our approach generates constraints and invalid values, we did not use this in this study. Moving forward, we plan to create a chatbot for tester interaction to change generated values and obtain more accurate values from the model, thereby improving the effectiveness and coverage of filling web forms.

## References

[1] K. Wakil and D. N. A. Jawawi, "Intelligent web applications as future generation of web applications," Scientific Journal of Informatics, vol. 6, no. 2, p. 214, 2019.

[2] A. Hoffman, Web application security. O'Reilly Media, Inc., 2024.

[3] International Organization for Standardization (ISO) and International Electrotechnical Commission (IEC), "ISO/IEC 25010:2011, Systems and software engineering – Systems and software Quality Requirements and Evaluation (SQuaRE) – System and software quality models," International Organization for Standardization, Geneva, Switzerland, 2011, standard.

[4] A. K. Behera, P. Chaudhury, and C. S. K. Dash, "A comprehensive survey on intelligent software reliability prediction," Discover Computing, vol. 28, no. 1, p. 90, 2025.

[5] International Organization for Standardization and Institute of Electrical and Electronics Engineers, "Software and systems engineering – software testing – part 2: Test processes," ISO/IEC/IEEE, Geneva, Switzerland, Tech. Rep. 29119-2:2021, 2021.

[6] F. YazdaniBanafsheDaragh and S. Malek, "Deep gui: Black-box gui input generation with deep learning," in 2021 36th IEEE/ACM International Conference on Automated Software Engineering (ASE). IEEE, 2021.

[7] S. Nidhra and J. Dondeti, "Black box and white box testing techniques- a literature review," International Journal of Embedded Systems and Applications (IJESA), vol. 2, no. 2, pp. 29–50, 2012.

[8] P. Ammann and J. Offutt, Introduction to software testing. Cambridge University Press, 2017.

[9] J. Wang et al., "Software testing with large language models: Survey, landscape, and vision," IEEE Transactions on Software Engineering, 2024.

[10] S. Neel and P. Chang, "Privacy issues in large language models: A survey," arXiv preprint arXiv:2312.06717, 2023.

[11] N. Sunman, Y. Soydan, and H. Sözer, "Automated web application testing driven by pre-recorded test cases," Journal of Systems and Software, vol. 193, p. 111441, 2022.

[12] J.-W. Lin, F. Wang, and P. Chu, "Using semantic similarity in crawling-based web application testing," in 2017 IEEE International Conference on Software Testing, Verification and Validation (ICST). IEEE, 2017.

[13] A. Groce, "Coverage rewarded: Test input generation via adaptation-based programming," in 2011 26th IEEE/ACM International Conference on Automated Software Engineering (ASE). IEEE, 2011.

[14] S. Carino and J. H. Andrews, "Dynamically testing guis using ant colony optimization (t)," in 2015 30th IEEE/ACM International Conference on Automated Software Engineering (ASE). IEEE, 2015.

[15] Y. Zheng et al., "Automatic web testing using curiosity-driven reinforcement learning," in 2021 IEEE/ACM 43rd International Conference on Software Engineering (ICSE). IEEE, 2021.

[16] D. P. Nguyen and S. Maag, "Codeless web testing using selenium and machine learning," in ICSOFT 2020: 15th International Conference on Software Technologies. ScitePress, 2020.

[17] Z. Liu et al., "Fill in the blank: Context-aware automated text input generation for mobile gui testing," in 2023 IEEE/ACM 45th International Conference on Software Engineering (ICSE). IEEE, 2023.

[18] F.-K. Chen, C.-H. Liu, and S. D. You, "Using large language model to fill in web forms to support automated web application testing," Information, vol. 16, no. 2, p. 102, 2025.

[19] C. Raffel, N. Shazeer, A. Roberts, K. Lee, S. Narang, M. Matena, Y. Zhou, W. Li, and P. J. Liu, "Exploring the limits of transfer learning with a unified text-to-text transformer," Journal of Machine Learning Research, vol. 21, no. 140, pp. 1–67, 2020. [Online]. Available: http://jmlr.org/papers/v21/20-074.html

[20] D. Bram, "Mocker-data-generator," https://github.com/danibram/mocker-data-generator, 2024.

[21] N.-K. Le et al., "Automated web application testing: End-to-end test case generation with large language models and screen transition graphs," arXiv preprint arXiv:2506.02529, 2025.

[22] T. Li et al., "Large language models for automated web-form-test generation: An empirical study," ACM Transactions on Software Engineering and Methodology, 2025.

[23] A. Dubey, A. Jauhri, A. Pandey, A. Kadian, A. Al-Dahle, A. Letman, A. Mathur, A. Schelten, A. Yang, A. Fan et al., "The llama 3 herd of models," arXiv e-prints, pp. arXiv–2407, 2024.

[24] "Timeoff.management," https://github.com/timeoff-management/application, gitHub repository, Accessed: 2025-07-30.

[25] "Nodebb," https://github.com/NodeBB/NodeBB, gitHub repository, Accessed: 2025-07-30.
[26] "Keystonejs," https://github.com/keystonejs/keystone, gitHub repository, Accessed: 2025-07-30.
[27] "Django blog," https://github.com/reljicd/django-blog, gitHub repository, Accessed: 2025-07-30.
[28] "Spring petclinic," https://github.com/spring-projects/spring-petclinic, gitHub repository, Accessed: 2025-07-30.
[29] Sherin, Salman, et al. "QExplore: An exploration strategy for dynamic web applications using guided search." Journal of Systems and Software 195 (2023): 111512.
[30] "Hello GPT-4o." https://openai.com/index/hello-gpt-4o/ Accessed: 2025-07-30.
[31] "Divar," http://divar.ir/, accessed: 2025-07-30.
[32] "Zoomit," https://www.zoomit.ir/, accessed: 2025-07-30.
[33] "Blog.ir," https://blog.ir/, accessed: 2025-07-30.
[34] "Blogfa," https://blogfa.com/, accessed: 2025-07-30.
[35] "Soft98," https://soft98.ir/, accessed: 2025-07-30.
[36] "Digikala," https://www.digikala.com/, accessed: 2025-07-30.
[37] "Ninisite," https://www.ninisite.com/, accessed: 2025-07-30.
[38] "Shatel user panel," https://my.shatel.ir/, accessed: 2025-07-30.
[39] "Bmi baam," https://baam.bmi.ir/, accessed: 2025-07-30.
[40] "FUM Portal," http://pooya.um.ac.ir/, accessed: 2025-07-30.